\documentclass[sn-mathphys-num]{sn-jnl}

\usepackage{graphicx}%
\usepackage{multirow}%
\usepackage{amsmath,amssymb,amsfonts}%
\usepackage{amsthm}%
\usepackage{mathrsfs}%
\usepackage[title]{appendix}%
\usepackage{xcolor}%
\usepackage{textcomp}%
\usepackage{manyfoot}%
\usepackage{booktabs}%
\usepackage{algorithm}%
\usepackage{algorithmicx}%
\usepackage{algpseudocode}%
\usepackage{listings}%

\begin{document}

\title[Article Title]{Sensing Few Electrons Floating on Helium with High-Electron-Mobility Transistors}


\author*[1,2]{\fnm{M. M.} \sur{Feldman}}\equalcont{These authors contributed equally to this work.}\email{mayerf@princeton.edu}

\author*[1]{\fnm{G.} \sur{Fuchs}}
\equalcont{These authors contributed equally to this work.}\email{gordian.fuchs@princeton.edu}

\author[1]{\fnm{T.} \sur{Liu}}
\equalcont{These authors contributed equally to this work.}

\author[3] {\fnm{L. A.} \sur{D'Imperio}}

\author[3] {\fnm{ M. D.} \sur{Henry}}

\author[3] {\fnm{E. A.} \sur{Shaner}}

\author*[1,4]{\fnm{S. A.} \sur{Lyon}}\email{lyon@princeton.edu}

\affil[1]{\orgdiv{Department of Electrical and Computer Engineering}, \orgname{Princeton University}, \orgaddress{\street{41 Olden Street}, \city{Princeton}, \postcode{08544}, \state{New Jersey}, \country{USA}}}

\affil[2]{\orgdiv{Department of Physics}, \orgname{Princeton University}, \orgaddress{\street{Washington Road}, \city{Princeton}, \postcode{08544}, \state{New Jersey}, \country{USA}}}

\affil[3]{\orgname{Microsystems Engineering, Science and Applications, Sandia National Laboratories}, \orgaddress{\city{Albuquerque}, \state{New Mexico}, \postcode{87123}, \country{USA}}}

\affil[4]{\orgname{EeroQ Corporation}, \orgaddress{\city{Chicago}, \state{Illinois}, \country{USA}}}


\abstract{We report on low-frequency measurements of few electrons floating on superfluid helium using a bespoke cryogenic cascode amplifier circuit built with off-the-shelf GaAs High-Electron-Mobility Transistors (HEMTs). We integrate this circuit with a Charge-Coupled Device (CCD) to transport the electrons on helium and characterize its performance. We show that this circuit has a Signal-to-Noise ratio (SNR) of $\thicksim$ 2$\frac{e}{\sqrt{Hz}}$ at 102 kHz, an order of magnitude improvement from previous implementations and provides a compelling alternative to few electron sensing with high frequency resonators.}




\maketitle

\section{Introduction}\label{sec1}

Electrons on helium has been shown to be a pristine platform for exploring fundamental condensed matter systems \cite{Ortiz1999,Spivak2010,Grimes1979,Andrei1997}. Due to its isolated nature, the electron spin floating on helium has also been proposed as the building block of a quantum computer \cite{Lyon2006}. However, probing the spin-degree of freedom of individual electrons in this platform requires precise control and detection of the electron number. Superconducting resonators \cite{Koolstra2019} and superconducting Single-Electron-Transistors \cite{Papageorgiou2005} (SETs) have led the charge on this front, reliably sensing single electrons. 

Commercial Heterojunction Bipolar Transistors \cite{Elarabi2021} (HBTs) and High-Electron Mobility Transistors \cite{Sabouret2008,Bradbury2011,Takita_2014,Belianchikov2024} (HEMTs) have been studied as replacements for the more exotic sensors but have yet to show overall single digit electron sensitivities, either due to excessive parasitic circuit capacitances \cite{Sabouret2008,Bradbury2011,Takita_2014} or weak single electron coupling to the underlying gate electrodes \cite{Elarabi2021,Belianchikov2024}. This is in stark contrast to implementations in the semiconductor quantum dot community \cite{Vink2007,Curry2015,Tracy2016,Curry2019,Blumhoff2022,Mills2022}, in which direct ohmic contacts and the use of amplifying proximal sensor dots has resulted in fault-tolerant qubit readout metrics \cite{Blumhoff2022,Mills2022}. Our new HEMT-based cascode amplifier described in this article takes a meaningful step towards this goal, achieving sensitivities of 2$\frac{e}{\sqrt{Hz}}$ at 102 kHz in typical channel device CCD architectures \cite{Boyle1970,Sabouret2008,Bradbury2011,Takita_2014,Marty1986}, an improvement of an order of magnitude from 15$\frac{e}{\sqrt{Hz}}$ shown previously \cite{Bradbury2010}.

\section{Device Overview}\label{sec2}

\begin{figure}
\centering
\includegraphics[width=1\textwidth]{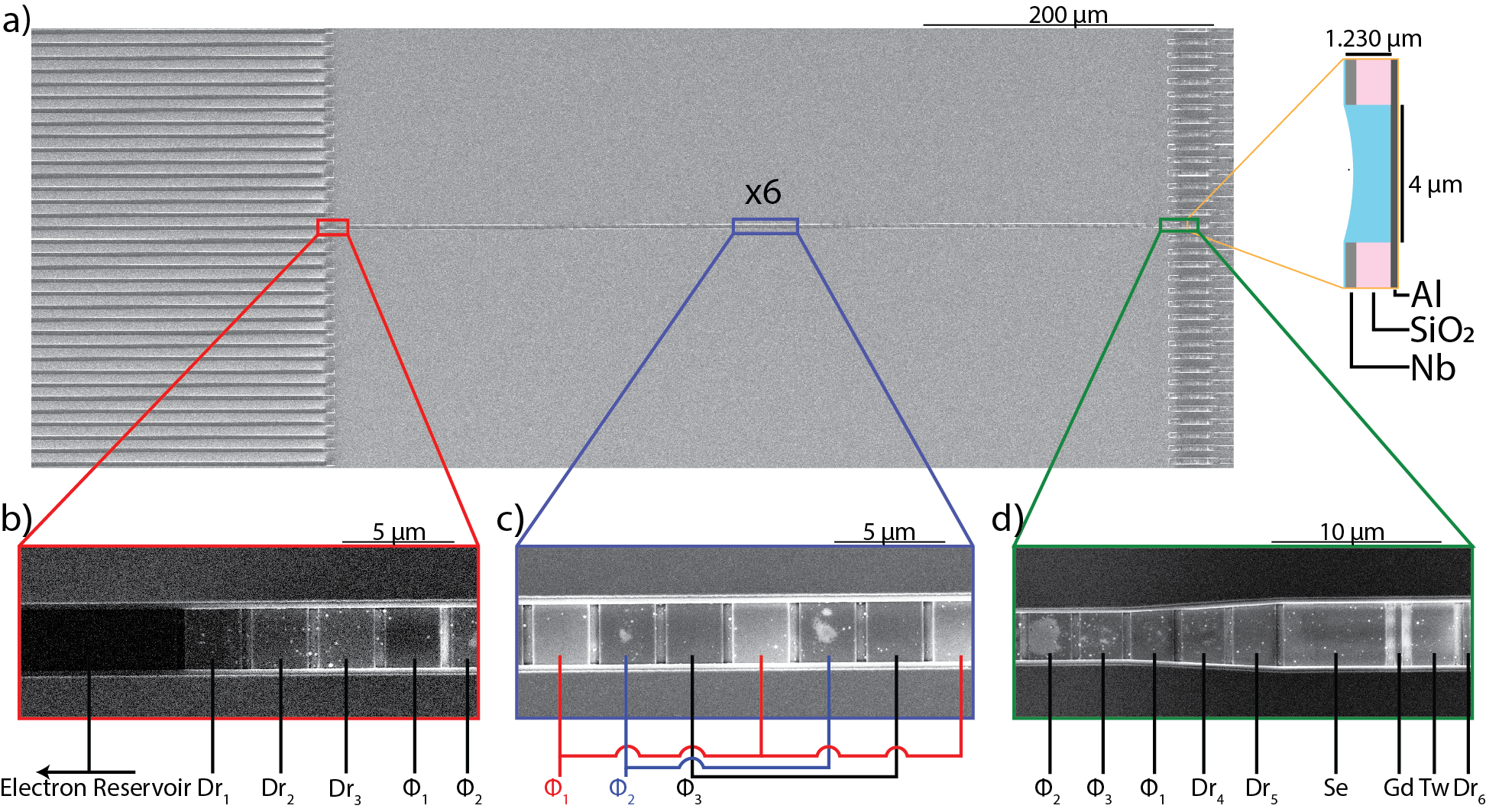}
\caption{SEM images of a device identical to the one used in the experiment and a cartoon mock-up of the channel profile. (a) Wide-field view of one of the six CCD arrays connecting the large electron reservoir (left) to the Twiddle-Sense region (right). (b-d) Zoomed in SEM images of each region with gate labels. The gates in the central part, (c), are connected as a 3-phase CCD where each CCD phase is labeled as $\phi_i$. Each region is segmented by separately controlled Door gates labeled $\mathrm{Dr}_i$. Gates with identical labels are controlled with the same voltage source. Note in (d) the channel widens from 3 $\mu$m to 4 $\mu$m to reduce screening of underlying gates by the top gate, while $\mathrm{Dr}_6$ is always kept negative to keep electrons from other parts of the device that are not relevant to  this experiment.}\label{Device_SEM_Image}
\end{figure}

The CCD device is made using Sandia National Laboratories MESA fab’s complementary metal-oxide-semiconductor (CMOS) back end of line (BEOL) metallization process.  Devices were fabricated at the 8” wafer scale using AlCu routing metal and planarized tungsten vias, where the top two layers of metal define the channel and gate geometry. The top layer of metal is made of $\sim$250 nm thick sputtered niobium, which is superconducting at this experimental temperature but not relevant in the experiment described here, while the submerged gates are made of $\sim$900 nm thick aluminum. Shown in Fig. \ref{Device_SEM_Image}(a), the device is split into three regions, the electron reservoir, the 6 charge-shuttling channels, and the sensing region. The electron reservoir is made up of 486 parallel channels with three gates underneath that allow for capturing electrons and measuring electron density \cite{Sommer1971}. To better control and measure few electrons, only 6 shuttling channels were opened using an ICP-RIE based etching process. The transition from one region to the next is defined by at least two Door gates, allowing for controlled loading and unloading of electrons between regions. Close-ups of these transition regions are shown in Fig. \ref{Device_SEM_Image}(b-d) and all subsequent gate electrodes are labeled. All channels are 1.230 $\mu$m deep and 3 $\mu$m wide except in the Twiddle-Sense region, where the channel is tapered to be 4 $\mu$m wide to allow for better capacitive coupling between the Sense gate and the electrons floating above it. A cartoon of the vertical linecut of the Twiddle-Sense channel is shown to the right of Fig. \ref{Device_SEM_Image}(a). Given the channel height of 1.230 $\mu$m, a gate separation of 0.5 $\mu$m ensures seamless electron transport between gate electrodes. The Guard (Gd) gate is 0.5 $\mu$m wide to decrease the parasitic capacitance between the Sense (Se) and Twiddle (Tw) gates, while the Sense gate is 6 $\mu$m to increase the capacitive coupling to electrons floating above. The Twiddle, Door and CCD gates are all 2.5 $\mu$m wide.

\section{Cryogenic Amplifier}\label{sec3}

\begin{figure}[t]
\centering
\includegraphics[width=1\textwidth]{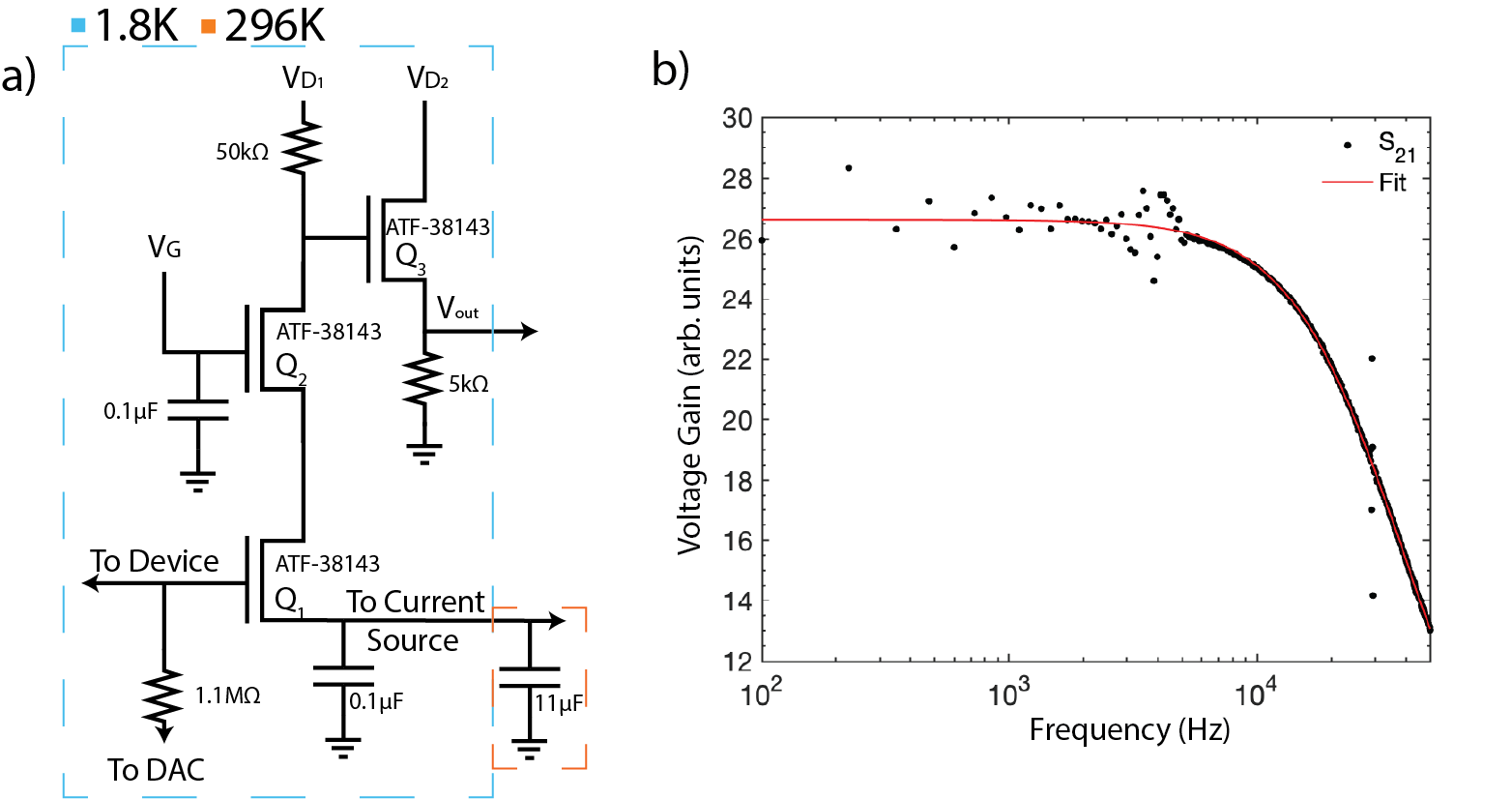}
\caption{(a) Schematic of the cryogenic amplifier circuit showing component values, biasing scheme, and DC coupling to device. (b) The roll-off frequency of the amplifier circuit when driven through the 1.1 M$\Omega$ resistor. This frequency is used to estimate the parasitic capacitance of the circuit and the expected measured voltage signal from a single electron.}\label{amp}
\end{figure}

A schematic of the cryogenic amplifier circuit is shown in Fig. \ref{amp}(a). This circuit is placed on the backside of the Printed Circuit Board (PCB) holding the device to ensure that no electrons are emitted onto the electrical components and placed as close as possible to the bondpads to minimize parasitic capacitance. The HEMTs are pseudomorphic depletion mode devices (ATF-38143) made by Avago whose bias voltages are tuned using separate channels of a home-built 18-bit Digital-to-Analog converter (DAC). We supply the source of the input HEMT with 15 $\mu$A of current via a capacitively-bypassed current source, ensuring low first-stage power dissipation. According to simulations, tens of picoFarad of capacitance to ground between the first stage and cascode HEMT will not degrade the performance of the amplifier circuit, allowing for the placement of the cascode and source-follower components at higher temperature stages that can support the added power dissipation.  The voltage measured at the input of the source determines the voltages of the rest of the circuit. We use a current source bypassed by a large capacitance to alleviate the burden of tuning the amplifier circuit at cryogenic temperatures which can vary drastically from the optimal room temperature parameters. This can be attributed to device specific temperature variability as well as large temperature variability in resistor values when cooling below 4 Kelvin. Typical operating voltages on the HEMTs are: $V_G = 0.6$ V, $V_{D_1} = 2.35$ V, and $V_{D_2} = 1.985$ V. The input is held near V = 0 V through the 1.1 M$\Omega$ (at low temperature) resistor.

\begin{figure}[t]
\centering
\includegraphics[width=1\textwidth]{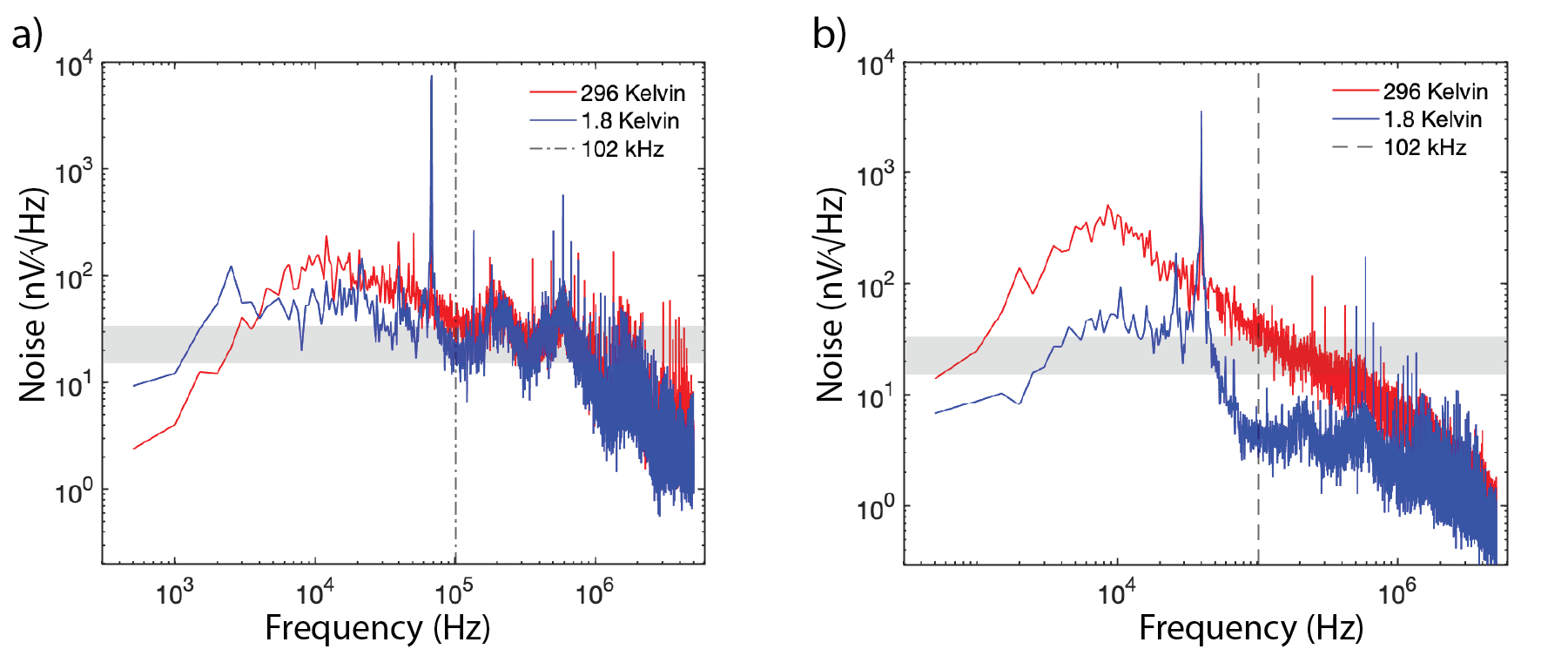}
\caption{The power spectral density of the cryogenic amplifier circuit with (a) and without (b) a device attached to the input. The shaded grey box is the expected signal from a single electron for 1 second of integration time for typical parasitic capacitances of our circuits and boards. A dotted line at 102 kHz serves as a visual reminder of our measurement frequency.}\label{noise_spectra}
\end{figure}

Fig. \ref{amp}(b) and Fig. \ref{noise_spectra} represent the figures of merit for the amplifier. The ATF-38143s quote an input capacitance of 1.2 pF. If the largest capacitance of the circuit is 1 pF, the voltage induced by a single electron is governed by a simple $Q=CV$ relation, which would yield 160 $\frac{nV}{e}$. We measure the transmitted signal (S$_{21}$) through the amplifier circuit by feeding a 1 mV$_{pp}$ signal into the 1.1 M$\Omega$ bias resistor and monitoring the output signal of the amplifier with a low-frequency network analyzer (HP 3577A). The input capacitance can be extracted by calculating the RC time constant of the measured amplifier roll-off with R = 1.1 M$\Omega$. 
A typical roll-off curve can be seen in Fig. \ref{amp}(b), which indicates $C_{p} \approx 5.25$ pF. According to simulations, the cascode transistor (Q2) eliminates most of the Miller enhancement of source-drain capacitance of Q1. We attribute a non-negligible portion of the capacitance to be related to the layout of the PCB. Simple fixes such as minimizing trace lengths and removing ground planes can yield a decrease in parasitic capacitance by more than a factor of 2. Fig. \ref{noise_spectra}(a-b) show the noise power-spectral density referenced to the input of the HEMT (Q1) with (a) and without (b) a device connected. The gray bars in each plot represent the range in which we expect the signal from a single electron to be if integrated for 1 second. Without a device connected, we find that we have a single electron SNR of 3 for a 1 Hz measurement bandwidth, while connecting a device yields an SNR of 0.5 for a 1 Hz measurement bandwidth. Since the noise does not decrease as a function of temperature, we conclude that most of it originates from room-temperature electronics and not the device itself. Modest room-temperature filtering is applied to the DC lines for gates around the Sense gate, although we suspect that adding cryogenic filtering would further improve our SNR.

\begin{figure}
\centering
\includegraphics[width=0.6\textwidth]{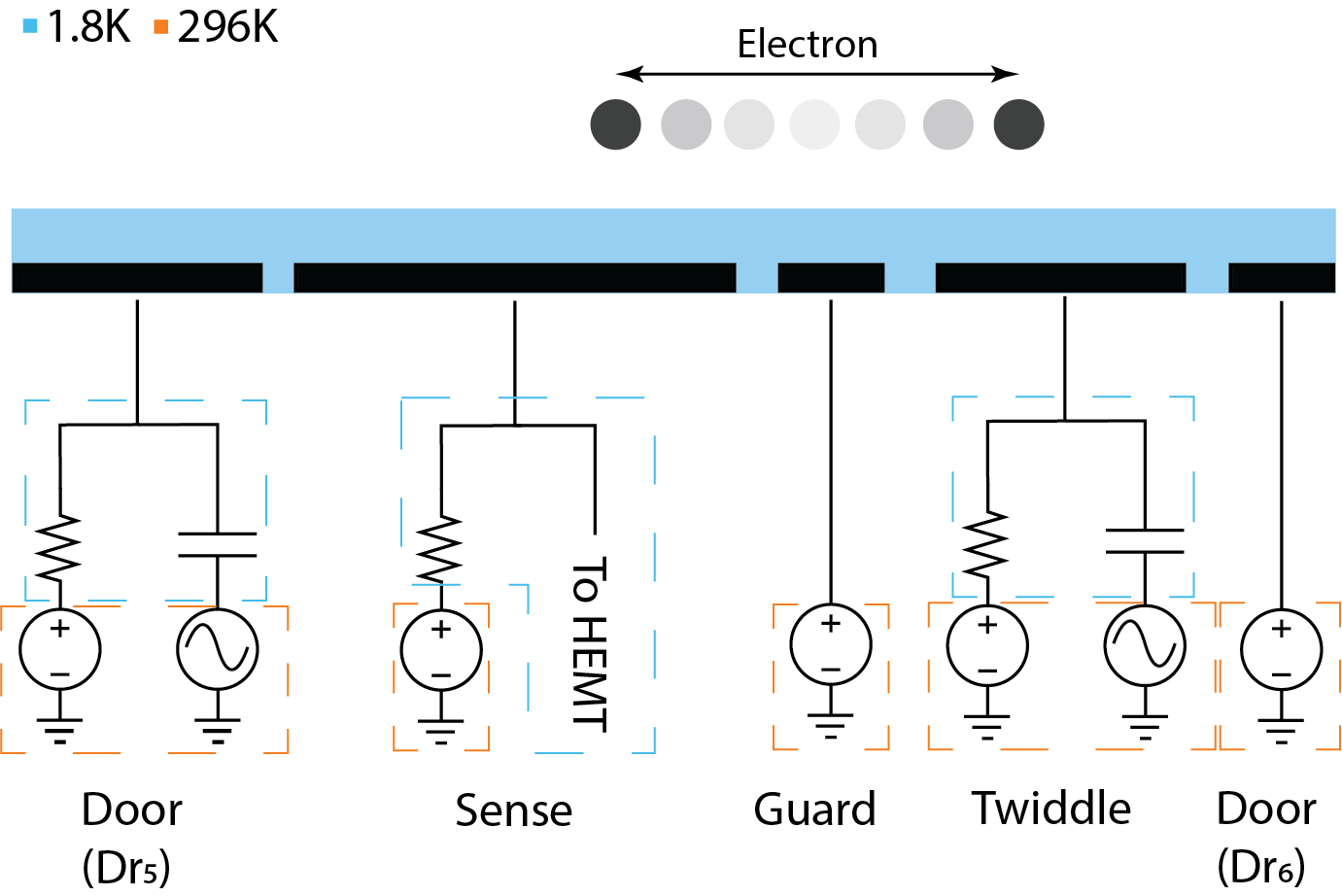}
\caption{A cartoon of the Twiddle-Sense measurement scheme where a single electron (indicated in black) oscillates back and forth across the Twiddle, Sense, and Guard gates. The difference in induced charge over the Sense gate is picked up by the on-board cryogenic amplifier circuit shown in Fig.\ref{amp}. An on-board bias-tee allows for the application of AC and DC signals on gates. The doors on either side of the region are biased negatively to ensure the electron(s) stay confined in the Twiddle-Sense region. The labels of these doors in parenthesis tie into the labels shown in Fig. \ref{Device_SEM_Image}.}\label{twiddle_cartoon}
\end{figure}

A schematic of the Twiddle-Sense measurement method is shown in Fig. \ref{twiddle_cartoon}(b). This method has been introduced previously to measure electrons in arrays of 78 and 120 channels \cite{Bradbury2011,Takita_2014}. The region consists of 5 gates, 2 of which are doors that are set to a negative voltage to corral electrons to the center. A 102 kHz, 300 mVpp sine-wave is fed into the Twiddle gate via a cryogenic bias-tee (R = 5 k$\Omega$ and C = 0.1 $\mu$F) while the DC voltage is kept at 0 V. The Guard gate acts both as a way to reduce parasitic capacitances between the Twiddle and Sense gates and to cut off electron transport between the two gates. Although this parasitic capacitance is greatly reduced, residual signal still bleeds into the Se gate. This is nulled by applying a sine-wave of opposite phase and smaller amplitude on the nearest door (Door$_5$ in Fig.1(d)). As the electron is pushed on and off the Sense gate, a signal is induced via its image charge which is then buffered by the cryogenic amplifier and measured with an SR830 lock-in amplifier. It is important to note that phase slips between the drive and nulling signal can result in extraneous signals that may spoil the analysis of the measurement scheme. To avoid this possibility, the drive and null are fed from the same two-channel AWG (Agilent 33622A) whose 10 MHz clock is supplied by a stable FS725 rubidium clock. No major variations in phase are observed during an experimental cycle. 

\section{Few Electron Sensing}\label{sec4}

Electrons are populated on the device by firing a tungsten filament located $\sim$2 mm above the device. We ensure that the majority of electrons are confined in the electron reservoir by setting all gates to -3 V while setting the electron reservoir gates to +2 V and achieve electron densities of $\sim 1.5\times 10^{9}$ cm$^{-2}$. After initial emission of electrons from the filament, the Twiddle, Sense, Guard, and electron reservoir gates are ramped to 0 V while the top metal is generally set to between -1 to -0.75 V. Due to the large number of electrons that are produced from thermionic emission and the small bias between the filament and negative gate electrodes, a few electrons can be found on the top metal of the device which can muddle the results of our controlled loading via the electron reservoir. We clean excess electrons from the top metal by ramping the top metal from -1 V to -2 V and back to -1 V. As a result, some electrons are pushed into the Twiddle-Sense regions, which we empty by sequentially setting Twiddle, Guard, Sense, Door$_5$, and Door$_4$ to a negative voltage and pushing electrons onto $\phi_1$. To make sure electrons are also moved out of the Twiddle-Sense regions in parallel to the 6 open channels used in this experiment, Door$_4$ is set to -2 V, while $\phi_1$ is ramped to -1.5 V to eject electrons in the parallel channels back onto the top metal. We then clock our CCDs in the 6 open channels to the electron reservoir and repeat this cleaning procedure until we do not observe any electrons in our Twiddle-Sense regions from ramping the top metal to a more negative voltage bias.

\begin{figure}
\centering
\includegraphics[width=1\textwidth]{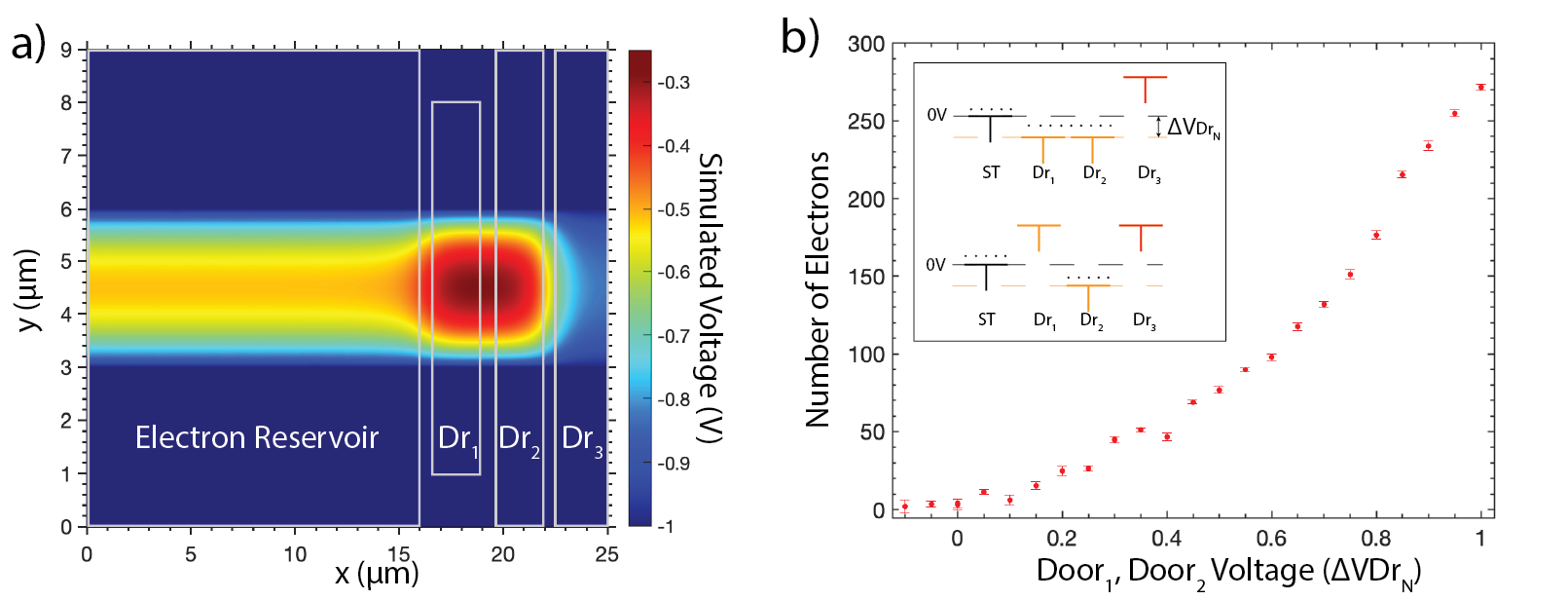}
\caption{(a) FEM simulation of the electric field profile around the electron reservoir door gates for a single channel. Gates are set to the following voltages: V$_{ST}$ = 0 V, V$_{Dr_1}$ = V$_{Dr_2}$ = +0.5 V, V$_{Dr_3}$ = -0.7 V, V$_{TM}$ = -1 V. The voltage at the position of the electron is approximately 47\% of what is set on the underlying gate electrodes. (b) Electron loading from the electron reservoir to the Twiddle-Sense gates as a function of the Door$_1$ and Door$_2$ voltages, with the electron reservoir held at 0 V.}\label{loading}
\end{figure}

To load electrons from the electron reservoir into Twiddle-Sense, we perform CCD clocking sequences while varying the Door$_1$ and Door$_2$ voltages. The two Door gates are initially ramped to the same voltage ($\Delta VDr_N$) to allow for electron diffusion. Door$_1$ is closed by setting it to -0.7 V, isolating electrons above Door$_2$ and allowing them to be clocked using the three-phase CCD gates ($\phi_1$, $\phi_2$, and $\phi_3$). The Door gates going into the Twiddle-Sense region (Door$_4$ and Door$_5$) are kept at +0.8 V to promote electron transfer from the last CCD gate ($\phi_1$). After the last clocking cycle is completed, Door$_4$ and Door$_5$ are sequentially ramped to -0.7 V and -2 V respectively and the Twiddle-Sense measurement is performed.
This measurement consists of two points, one in which the Guard is close to 0 V, allowing electrons to freely move on and off the Sense gate, and the other is when the Guard is set to -2 V, effectively cutting off the transport and the signal. Since the parasitic signal nulling is imperfect, the background signal at $V_{Guard}$ = -2 V is subtracted from both the real and imaginary components. The signal magnitude is then extracted by adding the two components in quadrature, and is analyzed using the following equation:

\begin{equation}
    n = \bigg(\frac{C_p V_{rms}}{e}\bigg)\bigg(\frac{2\sqrt{2}}{\alpha G}\bigg)
\end{equation}

where n is the number of electrons, $C_p$ is the parasitic capacitance measured from $S_{21}$ (Fig. \ref{amp}(b)), $V_{rms}$ is the rms voltage measured by the lock-in amplifier, $e$ is the electron charge, $\alpha$ is the fraction of electron charge that is induced on the Sense gate, and G is the amplifier gain. Using ANSYS Maxwell and the finite element method (FEM), Fig. \ref{loading}(a) shows the simulated electrostatic potential experienced by an electron with 0 V on the electron reservoir gate, +0.5 V on the two Door gates (Door$_1$ and Door$_2$), -0.7 V on Door$_3$, and -1 V on the top metal. We also solve for the electrostatic potential as we set a single gate to 1 V while keeping all other gates at 0 V and obtain an induced fractional electron charge of $\sim$0.47 for an electron floating directly above a single gate. The top metal still provides substantial screening in this configuration, augmenting the relative difference in potential that electrons floating at the top of the channel feel. A similar simulation not shown here is performed in the Twiddle-Sense region and reveals that although the widened channels reduce the top metal screening, the fractional electron charge induced on the Sense gate, $\alpha$, is $\sim$0.52.

Fig. \ref{loading}(b) shows the results of this experiment and the inset depicts the loading of electrons from the electron reservoir onto Door$_2$. Each data point is an average of 10 measurements at 1 second intervals with a time constant of 1 second on the lock-in amplifier. Between data points all electrons are returned to the reservoir before reloading. We find that above a certain loading voltage, electrons begin to move onto Door$_2$ from the electron reservoir, and can be shuttled into the Twiddle-Sense region. The numbers of electrons loaded increases with increasing Door gate voltages. With our experimental measurement parameters, we find an average noise of $\sim\pm$ 2 electrons. 

There are a few modifications that can be made to reach single electron sensitivity with this circuit.  Primary among them is the reduction of noise in the circuit. This can be done by placing cryogenic on-board low-pass filters on gates surrounding the Twiddle-Sense region. The experimental apparatus used here only contained room-temperature filters going from our DAC to a Breakout Box. To further improve the SNR, one can minimize the parasitic capacitance by placing components closer together which brings the circuit closer to the rated inter-electrode capacitance of the HEMTs. A new generation of PCBs have shown parasitic capacitances on the order of 2.5 pF. In the same vein, increasing the electron coupling, $\alpha$, by implementing wider channels \cite{Marty1986} or thin helium films \cite{Kleinbaum2018,Asfaw2019,Shaban2016} would result in an improvement of the SNR by $\sim$2. The HEMTs can also support higher frequency operations where $\frac{1}{f}$ noise dominates. By pushing to frequencies of 1-2 MHz after adding in-line low-pass filtering, we expect an additional improvement of one order of magnitude in SNR. Putting all of these improvements together can lead to a single electron SNR increase of two orders of magnitude, enabling faster and more reliable measurements in the future.

\section{Conclusion}\label{Conclusion}

In this experiment, we have shown the successful integration of a cryogenic amplifier based on commercially available HEMTs and its compatibility with low-frequency measurements. Their low input capacitance allows for an intrinsic SNR of 4 at a frequency of 102 kHz for single electron measurements. Although further optimization of room temperature equipment, filtering, PCB layout, and operation frequency is necessary to reach this SNR with a device connected, we have been able to reliably achieve an SNR of 1 for two electron detection using a device made via a commercially compatible BEOL CMOS processes which was confirmed over multiple cooldowns.

\bmhead{Acknowledgements}
We would like to thank Anthony Sigillito for the home-built Digital-To-Analog Converters, Kyle Castoria for useful discussions in regards to FEM analysis, and Bert Harrop for the use of his wire bonder.
This paper describes objective technical results and analysis. Any subjective views or opinions that might be expressed in the paper do not necessarily represent the views of the U.S. Department of Energy or the United States Government. Sandia National Laboratories is a multimission laboratory managed and operated by National Technology \& Engineering Solutions of Sandia, LLC, a wholly owned subsidiary of Honeywell International Inc., for the U.S. Department of Energy’s National Nuclear Security Administration under contract DE-NA0003525.
This material is primarily based upon work supported by the U.S. Department of Energy, Office of Science, Office of Basic Energy Sciences under Award $\#$ DE-SC0020136, as well as supported by the National Quantum Information Science Research Centers, Co-design Center for Quantum Advantage (C2QA) under Contract No. DE-SC0012704.
M.M.F was supported in part by the Program in Plasma Science and Technology at Princeton University. We acknowledge the use of Princeton’s Imaging and Analysis Center, which is partially supported by the Princeton Center for Complex Materials, a National Science Foundation MRSEC program (DMR-2011750).

\section*{Declarations}
S.A. Lyon is an officer of Eeroq Corp.

\bibliography{FewElectrons}


\begin{thebibliography}{25}
\ifx \bisbn   \undefined \def \bisbn  #1{ISBN #1}\fi
\ifx \binits  \undefined \def \binits#1{#1}\fi
\ifx \bauthor  \undefined \def \bauthor#1{#1}\fi
\ifx \batitle  \undefined \def \batitle#1{#1}\fi
\ifx \bjtitle  \undefined \def \bjtitle#1{#1}\fi
\ifx \bvolume  \undefined \def \bvolume#1{\textbf{#1}}\fi
\ifx \byear  \undefined \def \byear#1{#1}\fi
\ifx \bissue  \undefined \def \bissue#1{#1}\fi
\ifx \bfpage  \undefined \def \bfpage#1{#1}\fi
\ifx \blpage  \undefined \def \blpage #1{#1}\fi
\ifx \burl  \undefined \def \burl#1{\textsf{#1}}\fi
\ifx \doiurl  \undefined \def \doiurl#1{\url{https://doi.org/#1}}\fi
\ifx \betal  \undefined \def \betal{\textit{et al.}}\fi
\ifx \binstitute  \undefined \def \binstitute#1{#1}\fi
\ifx \binstitutionaled  \undefined \def \binstitutionaled#1{#1}\fi
\ifx \bctitle  \undefined \def \bctitle#1{#1}\fi
\ifx \beditor  \undefined \def \beditor#1{#1}\fi
\ifx \bpublisher  \undefined \def \bpublisher#1{#1}\fi
\ifx \bbtitle  \undefined \def \bbtitle#1{#1}\fi
\ifx \bedition  \undefined \def \bedition#1{#1}\fi
\ifx \bseriesno  \undefined \def \bseriesno#1{#1}\fi
\ifx \blocation  \undefined \def \blocation#1{#1}\fi
\ifx \bsertitle  \undefined \def \bsertitle#1{#1}\fi
\ifx \bsnm \undefined \def \bsnm#1{#1}\fi
\ifx \bsuffix \undefined \def \bsuffix#1{#1}\fi
\ifx \bparticle \undefined \def \bparticle#1{#1}\fi
\ifx \barticle \undefined \def \barticle#1{#1}\fi
\bibcommenthead
\ifx \bconfdate \undefined \def \bconfdate #1{#1}\fi
\ifx \botherref \undefined \def \botherref #1{#1}\fi
\ifx \url \undefined \def \url#1{\textsf{#1}}\fi
\ifx \bchapter \undefined \def \bchapter#1{#1}\fi
\ifx \bbook \undefined \def \bbook#1{#1}\fi
\ifx \bcomment \undefined \def \bcomment#1{#1}\fi
\ifx \oauthor \undefined \def \oauthor#1{#1}\fi
\ifx \citeauthoryear \undefined \def \citeauthoryear#1{#1}\fi
\ifx \endbibitem  \undefined \def \endbibitem {}\fi
\ifx \bconflocation  \undefined \def \bconflocation#1{#1}\fi
\ifx \arxivurl  \undefined \def \arxivurl#1{\textsf{#1}}\fi
\csname PreBibitemsHook\endcsname

\bibitem[\protect\citeauthoryear{Ortiz et~al.}{1999}]{Ortiz1999}
\begin{barticle}
\bauthor{\bsnm{Ortiz}, \binits{G.}},
\bauthor{\bsnm{Harris}, \binits{M.}},
\bauthor{\bsnm{Ballone}, \binits{P.}}:
\batitle{Zero temperature phases of the electron gas}.
\bjtitle{Phys. Rev. Lett.}
\bvolume{82},
\bfpage{5317}--\blpage{5320}
(\byear{1999})
\doiurl{10.1103/PhysRevLett.82.5317}
\end{barticle}
\endbibitem

\bibitem[\protect\citeauthoryear{Spivak et~al.}{2010}]{Spivak2010}
\begin{barticle}
\bauthor{\bsnm{Spivak}, \binits{B.}},
\bauthor{\bsnm{Kravchenko}, \binits{S.V.}},
\bauthor{\bsnm{Kivelson}, \binits{S.A.}},
\bauthor{\bsnm{Gao}, \binits{X.P.A.}}:
\batitle{Colloquium: Transport in strongly correlated two dimensional electron fluids}.
\bjtitle{Rev. Mod. Phys.}
\bvolume{82},
\bfpage{1743}--\blpage{1766}
(\byear{2010})
\doiurl{10.1103/RevModPhys.82.1743}
\end{barticle}
\endbibitem

\bibitem[\protect\citeauthoryear{Grimes and Adams}{1979}]{Grimes1979}
\begin{barticle}
\bauthor{\bsnm{Grimes}, \binits{C.C.}},
\bauthor{\bsnm{Adams}, \binits{G.}}:
\batitle{Evidence for a liquid-to-crystal phase transition in a classical, two-dimensional sheet of electrons}.
\bjtitle{Phys. Rev. Lett.}
\bvolume{42},
\bfpage{795}--\blpage{798}
(\byear{1979})
\doiurl{10.1103/PhysRevLett.42.795}
\end{barticle}
\endbibitem

\bibitem[\protect\citeauthoryear{Andrei}{1997}]{Andrei1997}
\begin{botherref}
\oauthor{\bsnm{Andrei}, \binits{E.Y.}}:
Two-Dimensional Electron System on Helium and other Cryogenic substrates.
Kulwar Academic Publishers, Dordrecht
(1997)
\end{botherref}
\endbibitem

\bibitem[\protect\citeauthoryear{Lyon}{2006}]{Lyon2006}
\begin{barticle}
\bauthor{\bsnm{Lyon}, \binits{S.A.}}:
\batitle{Spin-based quantum computing using electrons on liquid helium}.
\bjtitle{Phys. Rev. A}
\bvolume{74},
\bfpage{052338}
(\byear{2006})
\doiurl{10.1103/PhysRevA.74.052338}
\end{barticle}
\endbibitem

\bibitem[\protect\citeauthoryear{Koolstra et~al.}{2019}]{Koolstra2019}
\begin{barticle}
\bauthor{\bsnm{Koolstra}, \binits{G.}},
\bauthor{\bsnm{Yang}, \binits{G.}},
\bauthor{\bsnm{Schuster}, \binits{D.I.}}:
\batitle{Coupling a single electron on superfluid helium to a superconducting resonator}.
\bjtitle{Nature Communications}
\bvolume{10}(\bissue{1}),
\bfpage{5323}
(\byear{2019})
\doiurl{10.1038/s41467-019-13335-7}
\end{barticle}
\endbibitem

\bibitem[\protect\citeauthoryear{Papageorgiou et~al.}{2005}]{Papageorgiou2005}
\begin{barticle}
\bauthor{\bsnm{Papageorgiou}, \binits{G.}},
\bauthor{\bsnm{Glasson}, \binits{P.}},
\bauthor{\bsnm{Harrabi}, \binits{K.}},
\bauthor{\bsnm{Antonov}, \binits{V.}},
\bauthor{\bsnm{Collin}, \binits{E.}},
\bauthor{\bsnm{Fozooni}, \binits{P.}},
\bauthor{\bsnm{Frayne}, \binits{P.G.}},
\bauthor{\bsnm{Lea}, \binits{M.J.}},
\bauthor{\bsnm{Rees}, \binits{D.G.}},
\bauthor{\bsnm{Mukharsky}, \binits{Y.}}:
\batitle{Counting individual trapped electrons on liquid helium}.
\bjtitle{Applied Physics Letters}
\bvolume{86}(\bissue{15}),
\bfpage{153106}
(\byear{2005})
\doiurl{10.1063/1.1900301}
\end{barticle}
\endbibitem

\bibitem[\protect\citeauthoryear{Elarabi et~al.}{2021}]{Elarabi2021}
\begin{barticle}
\bauthor{\bsnm{Elarabi}, \binits{A.}},
\bauthor{\bsnm{Kawakami}, \binits{E.}},
\bauthor{\bsnm{Konstantinov}, \binits{D.}}:
\batitle{Cryogenic amplification of image-charge detection for readout of quantum states of electrons on liquid helium}.
\bjtitle{Journal of Low Temperature Physics}
\bvolume{202}(\bissue{5}),
\bfpage{456}--\blpage{465}
(\byear{2021})
\doiurl{10.1007/s10909-020-02552-w}
\end{barticle}
\endbibitem

\bibitem[\protect\citeauthoryear{Sabouret et~al.}{2008}]{Sabouret2008}
\begin{barticle}
\bauthor{\bsnm{Sabouret}, \binits{G.}},
\bauthor{\bsnm{Bradbury}, \binits{F.R.}},
\bauthor{\bsnm{Shankar}, \binits{S.}},
\bauthor{\bsnm{Bert}, \binits{J.A.}},
\bauthor{\bsnm{Lyon}, \binits{S.A.}}:
\batitle{Signal and charge transfer efficiency of few electrons clocked on microscopic superfluid helium channels}.
\bjtitle{Applied Physics Letters}
\bvolume{92}(\bissue{8}),
\bfpage{082104}
(\byear{2008})
\doiurl{10.1063/1.2884693}
\end{barticle}
\endbibitem

\bibitem[\protect\citeauthoryear{Bradbury et~al.}{2011}]{Bradbury2011}
\begin{barticle}
\bauthor{\bsnm{Bradbury}, \binits{F.R.}},
\bauthor{\bsnm{Takita}, \binits{M.}},
\bauthor{\bsnm{Gurrieri}, \binits{T.M.}},
\bauthor{\bsnm{Wilkel}, \binits{K.J.}},
\bauthor{\bsnm{Eng}, \binits{K.}},
\bauthor{\bsnm{Carroll}, \binits{M.S.}},
\bauthor{\bsnm{Lyon}, \binits{S.A.}}:
\batitle{Efficient clocked electron transfer on superfluid helium}.
\bjtitle{Phys. Rev. Lett.}
\bvolume{107},
\bfpage{266803}
(\byear{2011})
\doiurl{10.1103/PhysRevLett.107.266803}
\end{barticle}
\endbibitem

\bibitem[\protect\citeauthoryear{Takita and Lyon}{2014}]{Takita_2014}
\begin{barticle}
\bauthor{\bsnm{Takita}, \binits{M.}},
\bauthor{\bsnm{Lyon}, \binits{S.A.}}:
\batitle{Isolating electrons on superfluid helium}.
\bjtitle{Journal of Physics: Conference Series}
\bvolume{568}(\bissue{5}),
\bfpage{052034}
(\byear{2014})
\doiurl{10.1088/1742-6596/568/5/052034}
\end{barticle}
\endbibitem

\bibitem[\protect\citeauthoryear{Belianchikov et~al.}{2024}]{Belianchikov2024}
\begin{barticle}
\bauthor{\bsnm{Belianchikov}, \binits{M.}},
\bauthor{\bsnm{Kraus}, \binits{J.A.}},
\bauthor{\bsnm{Konstantinov}, \binits{D.}}:
\batitle{Cryogenic resonant amplifier for electron-on-helium image charge readout}.
\bjtitle{Journal of Low Temperature Physics}
\bvolume{215}(\bissue{5}),
\bfpage{312}--\blpage{323}
(\byear{2024})
\doiurl{10.1007/s10909-023-03033-6}
\end{barticle}
\endbibitem

\bibitem[\protect\citeauthoryear{Vink et~al.}{2007}]{Vink2007}
\begin{barticle}
\bauthor{\bsnm{Vink}, \binits{I.T.}},
\bauthor{\bsnm{Nooitgedagt}, \binits{T.}},
\bauthor{\bsnm{Schouten}, \binits{R.N.}},
\bauthor{\bsnm{Vandersypen}, \binits{L.M.K.}},
\bauthor{\bsnm{Wegscheider}, \binits{W.}}:
\batitle{Cryogenic amplifier for fast real-time detection of single-electron tunneling}.
\bjtitle{Applied Physics Letters}
\bvolume{91}(\bissue{12}),
\bfpage{123512}
(\byear{2007})
\doiurl{10.1063/1.2783265}
\end{barticle}
\endbibitem

\bibitem[\protect\citeauthoryear{Curry et~al.}{2015}]{Curry2015}
\begin{barticle}
\bauthor{\bsnm{Curry}, \binits{M.J.}},
\bauthor{\bsnm{England}, \binits{T.D.}},
\bauthor{\bsnm{Bishop}, \binits{N.C.}},
\bauthor{\bsnm{Ten-Eyck}, \binits{G.}},
\bauthor{\bsnm{Wendt}, \binits{J.R.}},
\bauthor{\bsnm{Pluym}, \binits{T.}},
\bauthor{\bsnm{Lilly}, \binits{M.P.}},
\bauthor{\bsnm{Carr}, \binits{S.M.}},
\bauthor{\bsnm{Carroll}, \binits{M.S.}}:
\batitle{Cryogenic preamplification of a single-electron-transistor using a silicon-germanium heterojunction-bipolar-transistor}.
\bjtitle{Applied Physics Letters}
\bvolume{106}(\bissue{20}),
\bfpage{203505}
(\byear{2015})
\doiurl{10.1063/1.4921308}
\end{barticle}
\endbibitem

\bibitem[\protect\citeauthoryear{Tracy et~al.}{2016}]{Tracy2016}
\begin{barticle}
\bauthor{\bsnm{Tracy}, \binits{L.A.}},
\bauthor{\bsnm{Luhman}, \binits{D.R.}},
\bauthor{\bsnm{Carr}, \binits{S.M.}},
\bauthor{\bsnm{Bishop}, \binits{N.C.}},
\bauthor{\bsnm{Ten~Eyck}, \binits{G.A.}},
\bauthor{\bsnm{Pluym}, \binits{T.}},
\bauthor{\bsnm{Wendt}, \binits{J.R.}},
\bauthor{\bsnm{Lilly}, \binits{M.P.}},
\bauthor{\bsnm{Carroll}, \binits{M.S.}}:
\batitle{Single shot spin readout using a cryogenic high-electron-mobility transistor amplifier at sub-kelvin temperatures}.
\bjtitle{Applied Physics Letters}
\bvolume{108}(\bissue{6}),
\bfpage{063101}
(\byear{2016})
\doiurl{10.1063/1.4941421}
\end{barticle}
\endbibitem

\bibitem[\protect\citeauthoryear{Curry et~al.}{2019}]{Curry2019}
\begin{barticle}
\bauthor{\bsnm{Curry}, \binits{M.J.}},
\bauthor{\bsnm{Rudolph}, \binits{M.}},
\bauthor{\bsnm{England}, \binits{T.D.}},
\bauthor{\bsnm{Mounce}, \binits{A.M.}},
\bauthor{\bsnm{Jock}, \binits{R.M.}},
\bauthor{\bsnm{Bureau-Oxton}, \binits{C.}},
\bauthor{\bsnm{Harvey-Collard}, \binits{P.}},
\bauthor{\bsnm{Sharma}, \binits{P.A.}},
\bauthor{\bsnm{Anderson}, \binits{J.M.}},
\bauthor{\bsnm{Campbell}, \binits{D.M.}},
\bauthor{\bsnm{Wendt}, \binits{J.R.}},
\bauthor{\bsnm{Ward}, \binits{D.R.}},
\bauthor{\bsnm{Carr}, \binits{S.M.}},
\bauthor{\bsnm{Lilly}, \binits{M.P.}},
\bauthor{\bsnm{Carroll}, \binits{M.S.}}:
\batitle{Single-shot readout performance of two heterojunction-bipolar-transistor amplification circuits at millikelvin temperatures}.
\bjtitle{Scientific Reports}
\bvolume{9}(\bissue{1}),
\bfpage{16976}
(\byear{2019})
\doiurl{10.1038/s41598-019-52868-1}
\end{barticle}
\endbibitem

\bibitem[\protect\citeauthoryear{Blumoff et~al.}{2022}]{Blumhoff2022}
\begin{barticle}
\bauthor{\bsnm{Blumoff}, \binits{J.Z.}},
\bauthor{\bsnm{Pan}, \binits{A.S.}},
\bauthor{\bsnm{Keating}, \binits{T.E.}},
\bauthor{\bsnm{Andrews}, \binits{R.W.}},
\bauthor{\bsnm{Barnes}, \binits{D.W.}},
\bauthor{\bsnm{Brecht}, \binits{T.L.}},
\bauthor{\bsnm{Croke}, \binits{E.T.}},
\bauthor{\bsnm{Euliss}, \binits{L.E.}},
\bauthor{\bsnm{Fast}, \binits{J.A.}},
\bauthor{\bsnm{Jackson}, \binits{C.A.C.}},
\bauthor{\bsnm{Jones}, \binits{A.M.}},
\bauthor{\bsnm{Kerckhoff}, \binits{J.}},
\bauthor{\bsnm{Lanza}, \binits{R.K.}},
\bauthor{\bsnm{Raach}, \binits{K.}},
\bauthor{\bsnm{Thomas}, \binits{B.J.}},
\bauthor{\bsnm{Velunta}, \binits{R.}},
\bauthor{\bsnm{Weinstein}, \binits{A.J.}},
\bauthor{\bsnm{Ladd}, \binits{T.D.}},
\bauthor{\bsnm{Eng}, \binits{K.}},
\bauthor{\bsnm{Borselli}, \binits{M.G.}},
\bauthor{\bsnm{Hunter}, \binits{A.T.}},
\bauthor{\bsnm{Rakher}, \binits{M.T.}}:
\batitle{Fast and high-fidelity state preparation and measurement in triple-quantum-dot spin qubits}.
\bjtitle{PRX Quantum}
\bvolume{3},
\bfpage{010352}
(\byear{2022})
\doiurl{10.1103/PRXQuantum.3.010352}
\end{barticle}
\endbibitem

\bibitem[\protect\citeauthoryear{Mills et~al.}{2022}]{Mills2022}
\begin{barticle}
\bauthor{\bsnm{Mills}, \binits{A.R.}},
\bauthor{\bsnm{Guinn}, \binits{C.R.}},
\bauthor{\bsnm{Feldman}, \binits{M.M.}},
\bauthor{\bsnm{Sigillito}, \binits{A.J.}},
\bauthor{\bsnm{Gullans}, \binits{M.J.}},
\bauthor{\bsnm{Rakher}, \binits{M.T.}},
\bauthor{\bsnm{Kerckhoff}, \binits{J.}},
\bauthor{\bsnm{Jackson}, \binits{C.A.C.}},
\bauthor{\bsnm{Petta}, \binits{J.R.}}:
\batitle{High-fidelity state preparation, quantum control, and readout of an isotopically enriched silicon spin qubit}.
\bjtitle{Phys. Rev. Appl.}
\bvolume{18},
\bfpage{064028}
(\byear{2022})
\doiurl{10.1103/PhysRevApplied.18.064028}
\end{barticle}
\endbibitem

\bibitem[\protect\citeauthoryear{Boyle and Smith}{1970}]{Boyle1970}
\begin{barticle}
\bauthor{\bsnm{Boyle}, \binits{W.S.}},
\bauthor{\bsnm{Smith}, \binits{G.E.}}:
\batitle{Charge coupled semiconductor devices}.
\bjtitle{Bell System Technical Journal}
\bvolume{49}(\bissue{4}),
\bfpage{587}--\blpage{593}
(\byear{1970})
\doiurl{10.1002/j.1538-7305.1970.tb01790.x}
\end{barticle}
\endbibitem

\bibitem[\protect\citeauthoryear{Marty}{1986}]{Marty1986}
\begin{barticle}
\bauthor{\bsnm{Marty}, \binits{D.}}:
\batitle{Stability of two-dimensional electrons on a fractionated helium surface}.
\bjtitle{Journal of Physics C: Solid State Physics}
\bvolume{19}(\bissue{30}),
\bfpage{6097}
(\byear{1986})
\doiurl{10.1088/0022-3719/19/30/019}
\end{barticle}
\endbibitem

\bibitem[\protect\citeauthoryear{Bradbury}{2010}]{Bradbury2010}
\begin{botherref}
\oauthor{\bsnm{Bradbury}, \binits{F.R.}}:
Cold electrons in siilcon and on superfluid helium.
PhD thesis,
Princeton University
(2010)
\end{botherref}
\endbibitem

\bibitem[\protect\citeauthoryear{Sommer and Tanner}{1971}]{Sommer1971}
\begin{barticle}
\bauthor{\bsnm{Sommer}, \binits{W.T.}},
\bauthor{\bsnm{Tanner}, \binits{D.J.}}:
\batitle{Mobility of electrons on the surface of liquid $^{4}\mathrm{He}$}.
\bjtitle{Phys. Rev. Lett.}
\bvolume{27},
\bfpage{1345}--\blpage{1349}
(\byear{1971})
\doiurl{10.1103/PhysRevLett.27.1345}
\end{barticle}
\endbibitem

\bibitem[\protect\citeauthoryear{Kleinbaum and Lyon}{2018}]{Kleinbaum2018}
\begin{barticle}
\bauthor{\bsnm{Kleinbaum}, \binits{E.I.}},
\bauthor{\bsnm{Lyon}, \binits{S.A.}}:
\batitle{Thermopower-based hot electron thermometry of helium surface states at 1.6 k}.
\bjtitle{Phys. Rev. Lett.}
\bvolume{121},
\bfpage{236801}
(\byear{2018})
\doiurl{10.1103/PhysRevLett.121.236801}
\end{barticle}
\endbibitem

\bibitem[\protect\citeauthoryear{Asfaw et~al.}{2019}]{Asfaw2019}
\begin{barticle}
\bauthor{\bsnm{Asfaw}, \binits{A.T.}},
\bauthor{\bsnm{Kleinbaum}, \binits{E.I.}},
\bauthor{\bsnm{Henry}, \binits{M.D.}},
\bauthor{\bsnm{Shaner}, \binits{E.A.}},
\bauthor{\bsnm{Lyon}, \binits{S.A.}}:
\batitle{Transport measurements of surface electrons in 200-nm-deep helium-filled microchannels above amorphous metallic electrodes}.
\bjtitle{Journal of Low Temperature Physics}
\bvolume{195}(\bissue{3}),
\bfpage{300}--\blpage{306}
(\byear{2019})
\doiurl{10.1007/s10909-018-02139-6}
\end{barticle}
\endbibitem

\bibitem[\protect\citeauthoryear{Shaban et~al.}{2016}]{Shaban2016}
\begin{barticle}
\bauthor{\bsnm{Shaban}, \binits{F.}},
\bauthor{\bsnm{Ashari}, \binits{M.}},
\bauthor{\bsnm{Lorenz}, \binits{T.}},
\bauthor{\bsnm{Rau}, \binits{R.}},
\bauthor{\bsnm{Scheer}, \binits{E.}},
\bauthor{\bsnm{Kono}, \binits{K.}},
\bauthor{\bsnm{Rees}, \binits{D.G.}},
\bauthor{\bsnm{Leiderer}, \binits{P.}}:
\batitle{The helium field effect transistor (ii): Gated transport of surface-state electrons through micro-constrictions}.
\bjtitle{Journal of Low Temperature Physics}
\bvolume{185}(\bissue{3}),
\bfpage{339}--\blpage{353}
(\byear{2016})
\doiurl{10.1007/s10909-016-1641-6}
\end{barticle}
\endbibitem

\end{thebibliography}

\end{document}